\documentclass[manuscript]{acmart}
\AtBeginDocument{%
  \providecommand\BibTeX{{%
    \normalfont B\kern-0.5em{\scshape i\kern-0.25em b}\kern-0.8em\TeX}}}

\copyrightyear{2021} 
\acmYear{2021} 
\setcopyright{acmcopyright}\acmConference[PEARC '21]{Practice and Experience in Advanced Research Computing}{July 19--22, 2021}{Virtual Conference}
\acmBooktitle{Practice and Experience in Advanced Research Computing (PEARC '21), July 19--22, 2021, Virtual Conference}
\acmPrice{15.00}
\acmDOI{10.1145/3311790.3400848}
\acmISBN{978-1-4503-6689-2/20/07}

\begin{document}

\title{A Multi-Protocol, Secure, and Dynamic Data Storage Integration Framework for Multi-tenanted Science Gateway Middleware}

\author{Dimuthu Wannipurage}
\email{dwannipu@iu.edu}
\affiliation{%
  \institution{Cyberinfrastructure Integration Research Center,}
 \institution{Indiana University}
  \city{Bloomington} 
  \state{IN}
  \country{USA}
   \postcode{47408}}

\author{Suresh Marru}
\email{smarru@iu.edu}
\affiliation{%
  \institution{Cyberinfrastructure Integration Research Center,}
 \institution{Indiana University}
  \city{Bloomington} 
  \state{IN}
    \country{USA}
   \postcode{47408}}

\author{Eroma Abeysinghe}
\email{eabeysin@iu.edu}
\affiliation{%
  \institution{Cyberinfrastructure Integration Research Center,}
 \institution{Indiana University}
  \city{Bloomington} 
  \state{IN}
    \country{USA}
   \postcode{47408}}

\author{Isuru Ranawaka}
\email{isjarana@iu.edu}
\affiliation{%
  \institution{Cyberinfrastructure Integration Research Center,}
 \institution{Indiana University}
  \city{Bloomington} 
  \state{IN}
    \country{USA}
   \postcode{47408}}
   
\author{Marcus Christie}
\email{machrist@iu.edu}
\affiliation{
  \institution{Cyberinfrastructure Integration Research Center,}
   \institution{Indiana University}
  \city{Bloomington} 
  \state{IN}
    \country{USA}
   \postcode{47408}}
   
\author{Marlon Pierce}
\email{marpierc@iu.edu}
\affiliation{%
  \institution{Cyberinfrastructure Integration Research Center,}
 \institution{Indiana University}
  \city{Bloomington} 
  \state{IN}
    \country{USA}
   \postcode{47408}}

\renewcommand{\shortauthors}{Wannipurage et al.}
\renewcommand{\shorttitle}{Apache Airavata Data Storage Framework}

\begin{abstract}
Science gateways are user-centric, end-to-end cyberinfrastructure for managing scientific data and executions of computational software on distributed resources. In order to simplify the creation and management of science gateways, we have pursued a multi-tenanted, platform-as-a-service approach that allows multiple gateway front-ends (portals) to be integrated with a consolidated middleware that manages the movement of data and the execution of workflows on multiple back-end scientific computing resources. An important challenge for this approach is to provide an end-to-end data movement and management solution that allows gateway users to integrate their own data stores with the gateway platform. These user-provided data stores may include commercial cloud-based object store systems, third-party data stores accessed through APIs such as REST endpoints, and users’ own local storage resources. In this paper, we present a solution design and implementation based on the integration of a managed file transfer (MFT) service (Airavata MFT) into the platform.  
\end{abstract}



\settopmatter{printfolios=true}
\maketitle

\section{Introduction}
Science Gateways \cite{sgci-survey} play a vital role in research cyberinfrastructure by bridging the gap between scientists, research computing, and data management infrastructure. They provide user-friendly, domain-specific user environments while absorbing complexities such as job schedulers, storage infrastructure management, scientific software, and security compliance that arise when using shared scientific computing resources. 

Mature software projects \cite{jalili2020galaxy, mclennan2010hubzero, hudak2018open, calegari2019web, airavata} exist to create science gateways. Science gateway platforms further simplify the process for creating and operating science gateways by offering hosted deployments, which may be multi-instance or multi-tenanted \cite{pierce2020integrating}. Apache Airavata \cite{airavata} is an open-source science gateway framework that powers the multi-tenanted Science Gateways Platform as a service (SciGaP) \cite{pierce2018supporting}.  Multiple science gateways can connect to a single, scalable deployment of SciGaP’s Apache Airavata-based middleware through its well-defined API. The Django Portal for Airavata \cite{christie2020extensible} provides a turnkey science gateway front end that can be used to create new tenants to the middleware.  A list of active SciGaP gateways and supported research fields is available from \cite{circ-collab}. 

In recent work, we have begun to decouple Apache Airavata into multiple autonomous components that interoperate to deliver comprehensive science gateway functionality, at the same time which can also act as standalone components. Major components of Apache Airavata include job and workflow management \cite{wannipurage2019implementing}, security \cite{ranawaka2020custos}, data management \cite{wannipurageopen}, and metadata management \cite{marru2015apache}.  Resources that must be integrated into a science gateway system may include multiple supercomputers and mass storage systems at various universities, resources owned by individual science gateway users on commercial clouds, and local resources directly operated by individual users.

In this paper, we investigate three current research problems in science gateway systems: 1) decoupling data movement from web and control traffic in science gateway cyberinfrastructure; 2) interfacing with heterogeneous storage systems that include POSIX file systems \cite{walli1995posix}, Object Stores \cite{factor2005object}, and Block Stores \cite{wu2010cloud}; and 3) supporting new transfer protocols. 

\begin{figure}[h]
  \centering
  \includegraphics[width=\linewidth]{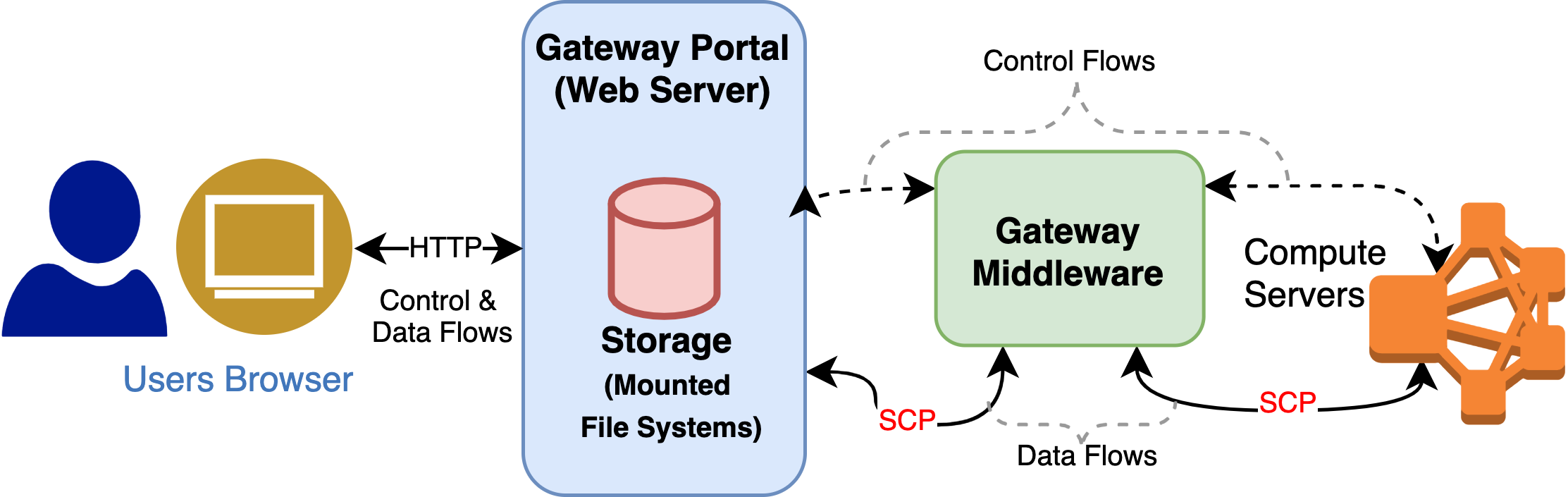}
  \caption{A conceptual view of canonical gateway architecture with four distinct layers. Data moves between users' local environments and remote computational clusters or storage devices through multiple hops of tiered services. These extra hops constrain data management flexibility, impede the design of highly available gateway systems, and add latencies and unnecessary network traffic.}
  \Description{A conceptual view of canonical gateway architecture with four distinct layers.}
  \label{figConceptualView}
\end{figure}

\section{Data Management Challenges in Science Gateways}

\textbf{Data Movement in Multi-Tiered Architectures:}
Science gateway cyberinfrastructure is typically conceptualized using a multi-tiered architecture \cite{pierce2018towards}, as illustrated in Fig. \ref{figConceptualView}. A web server hosts the gateway portal that interfaces with the user’s browser. The second tier bundles gateway middleware services such as API servers, state management and orchestration services, identity and security services, and persistent databases. The third tier consists of remote computing and data storage systems. The general challenge is the movement of data and management of permissions across tiers, where the browser (and its host computer), the portal server, the middleware, and the backend services are controlled by different groups.  

In a canonical deployment (Fig. \ref{figConceptualView}), transferring users' data to and from computational and storage resources is typically done by first copying to a large file system mounted to the web server. We will refer to this storage as the ``gateway portal data store.'' Having the data locally deployed with the gateway portal web server enables the portal to enforce data security with user-level authorization and serve the data to users through standard HTTP uploads and downloads. 

In a subsequent step, such as using the data as an input to an analysis or simulation, the middleware pulls the file from the gateway portal data store using standard protocols such as SCP and pushes the data stream to compute resources using SCP, SFTP, or similar protocols. The middleware may use in-memory buffering to optimize these transfers. This approach is rudimentary but provides a simple approach to developing end-to-end data transfer solutions. However, the extra and unneeded hop of transfers is inefficient and limits scalability of users and data.  

Science gateways execute scientific workflows and processing pipelines, so data are inputs and outputs of these processes. Gateways are responsible for fetching input data from users, submitting computational jobs to remote resources, and managing the outputs. Traditionally, users upload input data through gateway portals to the gateway portal data storage to make it available for the gateway middleware so that the middleware can use these inputs to submit jobs into remote computational resources. 

One of the main assumptions made in this scenario is that the gateway portal data storage and the portal server are deployed on the same server so that the portal can upload and download data by utilizing the local file system calls on top of the storage directory. Even though this is simple and easy to implement, this approach sacrifices the scalability of gateway storage as it is bounded by the maximum disk space available in the host machine or any directly attached network storage.

Another gateway use case is the capability of configuring and using cloud data storages as gateway storage. Cloud storage systems provide cheaper, easier to use, and significantly higher volume storage solutions for academic institutions, and as a result, researchers often store their research data in such cloud endpoints. When it comes to processing such data through a science gateway, it is cumbersome and inefficient to download and reupload that data into the gateway storage. A better solution is to  provide direct access from the gateway to these cloud endpoints rather than the typical POSIX-based storage acting as the broker. 

\begin{figure}[h]
  \centering
  \includegraphics[width=\linewidth]{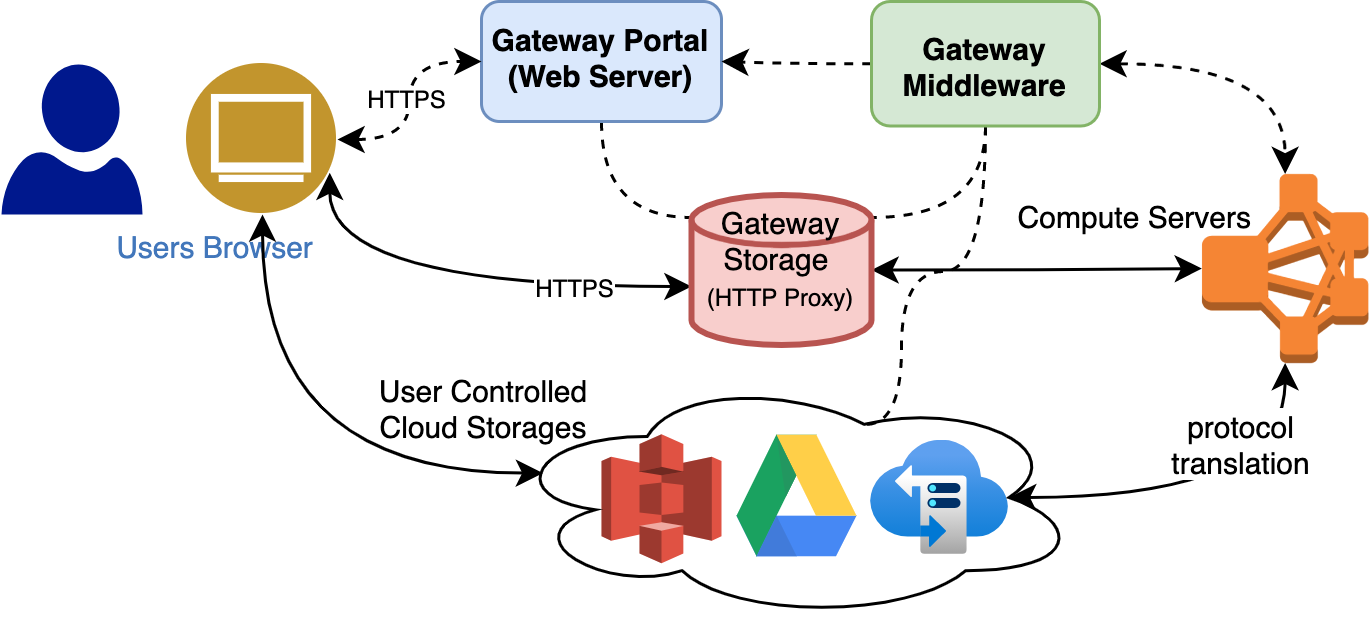}
  \caption{Decoupling the gateway storage from the webserver and serving by an HTTPS proxy eliminates redundant transfer hops while providing flexibility to add external storage without necessarily limiting web server deployments, high availability, and uptimes. The decoupled architecture also provides the capability to add user-controlled cloud storage but requires protocol translation and secured credential delegation. The dotted lines in the figure indicate control flow and solid lines data flow.}
  \Description{Decoupling the gateway storage from the webserver and serving by an HTTPS proxy eliminates redundant transfer hops while providing flexibility to add external storage without necessarily limiting web server deployments, high availability and uptimes. The decoupled architecture also provides the capability to add user controlled cloud storage but requires protocol translation and secured credential delegation. The dotted lines in the figure indicate control flow and solid lines data flow.}
  \label{figDecoupledStorage}
\end{figure}

We propose to address the above challenges by providing more flexibility in how science gateway cyberinfrastructure (Fig. \ref{figDecoupledStorage}) can mount storage endpoints and manage direct data transfer paths between storage endpoints and compute resources, eliminating data routing (but not control) through the gateway middleware.  

\textbf{An Extendable Managed File Transfer Framework for Science Gateways:}
To support the growing number of data transfer use cases, we adopted the distributed form of the ``separation of concerns'' design approach, collecting the data management and transfer functionality into a separate component, or service, to avoid introducing too much complexity into the middleware. This approach has the added potential benefit that the service, if properly designed, could work in standalone usage scenarios as well as in fully integrated science gateway scenarios. We propose a novel, extendable and multi protocol managed file transfer framework, Airavata MFT to address above mentioned aspects.

\begin{figure}[ht]
  \centering
  \includegraphics[width=\linewidth]{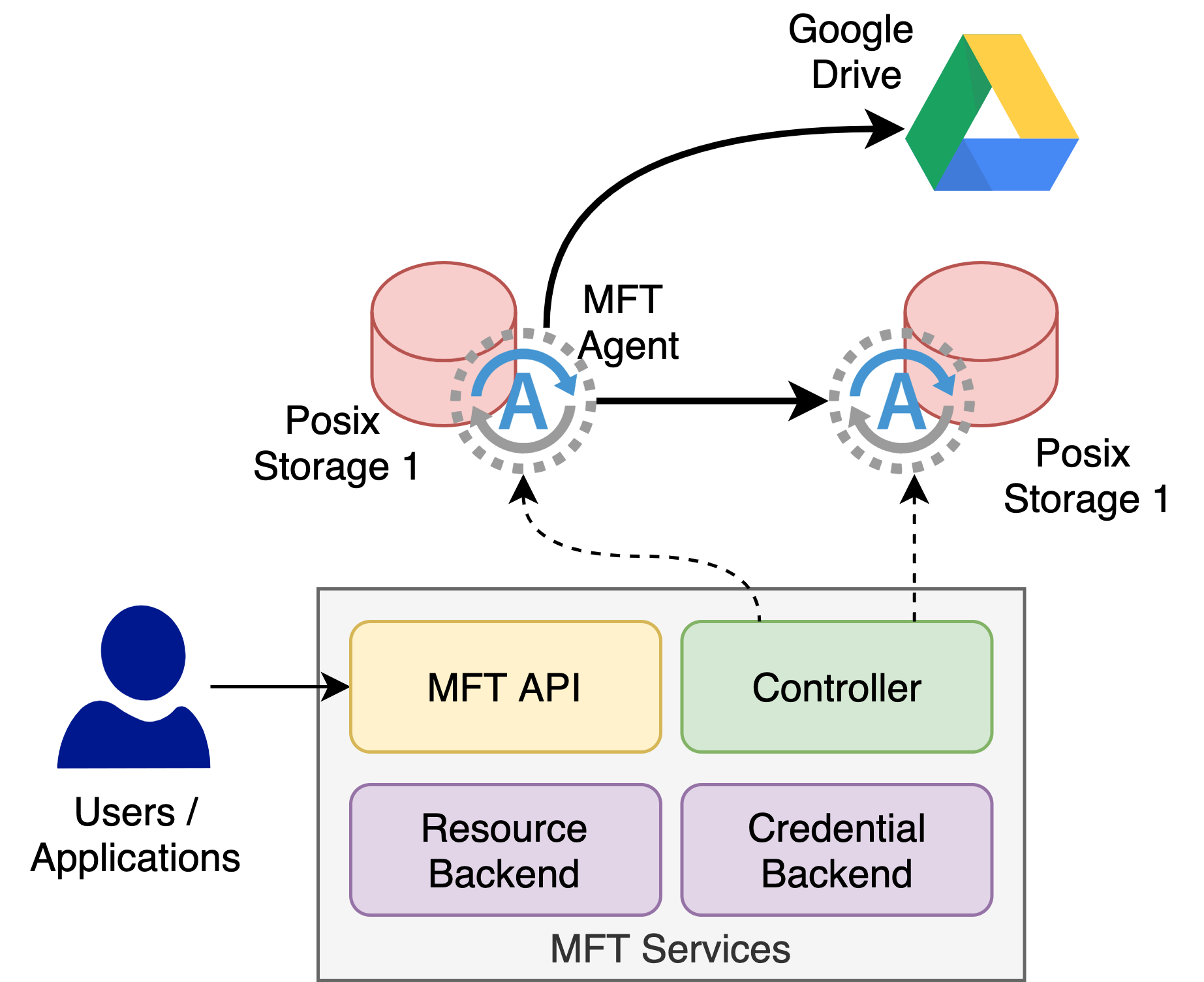}
  \caption{Top level Airavata MFT with two main data transfer modes; agent-to-agent transfer and agent-to-storage transfer. Thick solid arrows depict data transfer paths and dashed arrows depict transfer control paths.}
  \Description{Storage preferences.}
  \label{fig:mft-architecture}
\end{figure}

Airavata MFT has two main components: MFT Services and MFT Agents (Fig. \ref{fig:mft-architecture}). MFT Services are a set of microservices that work together to accept, monitor, and retry a transfer request coming into the framework. MFT Agents are the daemons installed in external storage systems that perform actual data transfers based on the instruction received by the MFT services. MFT Agents are installable binaries bundled with library implementations for protocols like SCP, SFTP, FTP, HTTP, and TUS. For cloud vendor systems such as Amazon S3, Google Drive, Azure Blobs, Box, and DropBox, these protocols and associated security mechanisms can be embedded as client libraries. These agents can be directly installed on target platforms. For example, an agent can be deployed on the same host as the storage or on a network level closer to the storage endpoint. These agents are capable of communicating with any storage endpoint type using supported protocols to either push or pull data using bundled protocol libraries. 

There are two modes of data transfer types in this approach: agent-to-agent and agent-to-storage transfers. In the agent-to-agent mechanism, the sending agent pulls file content from its local file system and pushes to the agent installed on the target storage. Agent-to-storage transfers have an agent on one end that communicates with the other storage endpoint using the protocol that the target storage endpoint supports. 

MFT Services consist of four main microservices; MFT API, Controller Service, Resource backend and Credential backend. The MFT API Service provides the front-facing application programming interface (API) for client applications to submit and monitor transfer requests. The Controller Service sends control instructions to external MFT Agents. Resource and Credential Backends are plugin interfaces to mount gateway-specific interfaces to fetch metadata and credentials information when performing a data transfer.

Airavata MFT is designed to separate data transfer paths and control paths at the architectural level. This provides flexible control of the data transfer route and allows us to implement end-to-end transfer quality of services like encryption and integrity verification. Fig~\ref{fig:mft-architecture} shows the command and transfer path separation of a general data transfer job handled by Airavata MFT. Users send data transfer requests to the MFT API, which forwards that message to the Controller. The Controller determines the corresponding agents that must be notified regarding the transfer. Resource and Credential Backends provide the connection details and credentials for those agents to connect securely to the other endpoints. Once those control messages are sent to the agent, the agent performs the transfer in a separate channel. 

\section{Integrating Airavata MFT into Science Gateways}

\begin{figure}[ht]
  \centering
  \includegraphics[width=\linewidth]{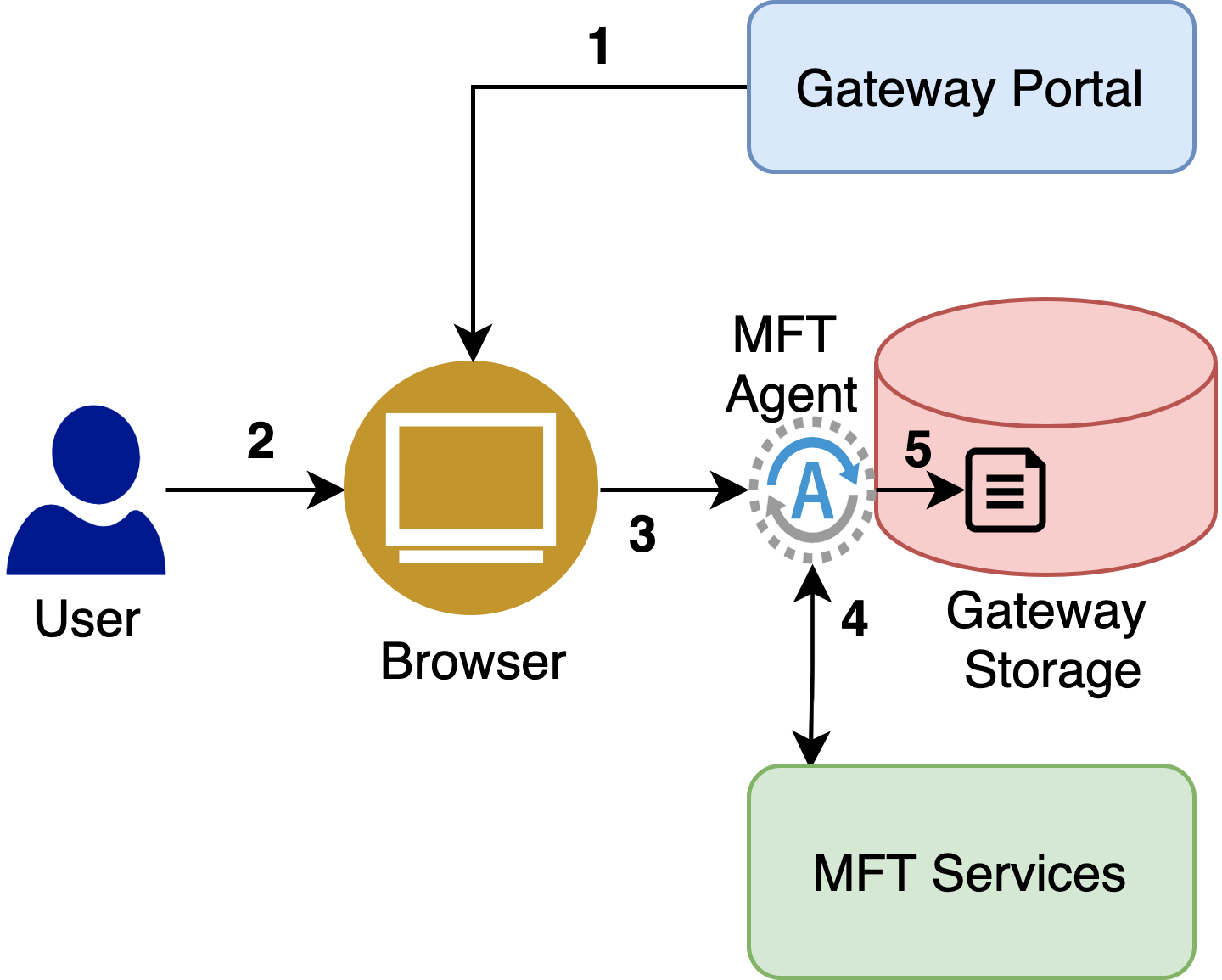}
  \caption{Airavata MFT uses agents and a coordinating service to extend gateway portal storage.
}
  \Description{group-based authorization}
  \label{figExtendedGatewayStorage}
\end{figure}

\textbf{Uploading and downloading data through the gateway portal:} Using Airavata MFT, we can logically and physically decouple science gateway portals from their local storage so that these can exist in two different instances. Fig. \ref{figExtendedGatewayStorage} shows the top level component organization of this approach with the workflow of actions in uploading or downloading data from the gateway storage. 

First, the gateway portal sends a file upload link in a hypermedia message to the user’s browser. This link points to the HTTP endpoint of the MFT Agent installed on the gateway storage. The user chooses files from his/her local storage and uploads these through the provided HTTP endpoint. This data stream is captured by the MFT Agent, which verifies the authenticity of the request including the user validation and the authorization of the user to access the gateway storage through MFT Services. Once the request is validated, the agent starts to deposit the file in the designated location of the gateway storage. 

\begin{figure}[ht]
  \centering
  \includegraphics[width=\linewidth]{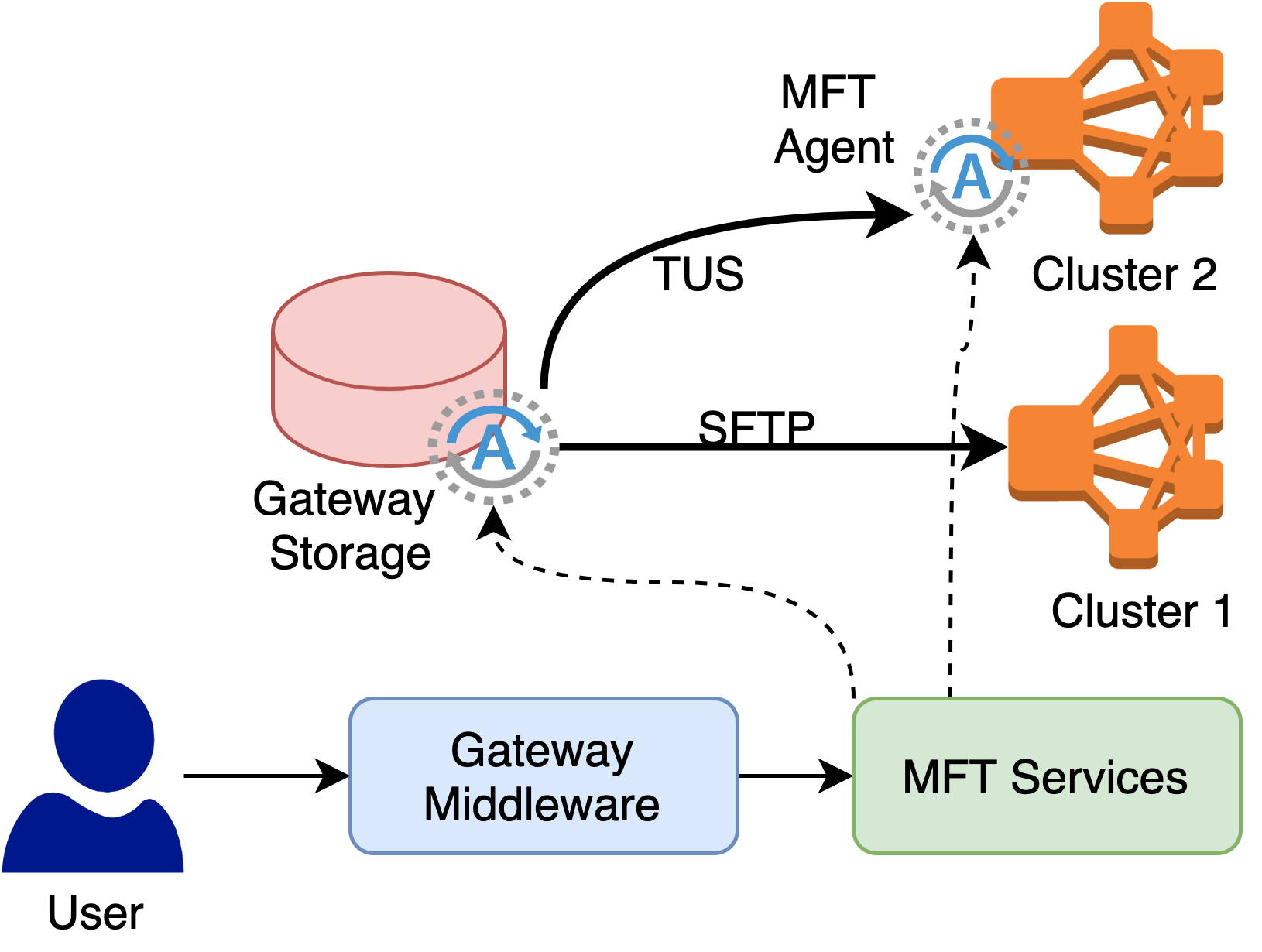}
  \caption{Implementing data transfer paths from gateway storage to the HPC cluster with Airavata MFT. Thick solid arrows depict data transfer paths and dashed arrows depict transfer control paths.
}
  \Description{Data Resource Management Service}
  \label{figStoragate2Cluster}
\end{figure}

\textbf{Data transfers between POSIX storage and HPC clusters:} Once the data is deposited in the gateways storage, the next step is to transfer the data to the HPC cluster as inputs for computational jobs. Airavata MFT separates the middleware, which controls the transfer, from the data transfer path, which is handled by the agents; see Fig. \ref{figStoragate2Cluster}.  If both endpoints have agents installed, then those agents open a HTTP2-based TUS \cite{tusSpec} data channel to transfer data in a secure and resumable manner. If only one endpoint has an agent installed, that agent uses the supported protocol on the other endpoint to transfer data. 

\begin{figure}[ht]
  \centering
  \includegraphics[width=\linewidth]{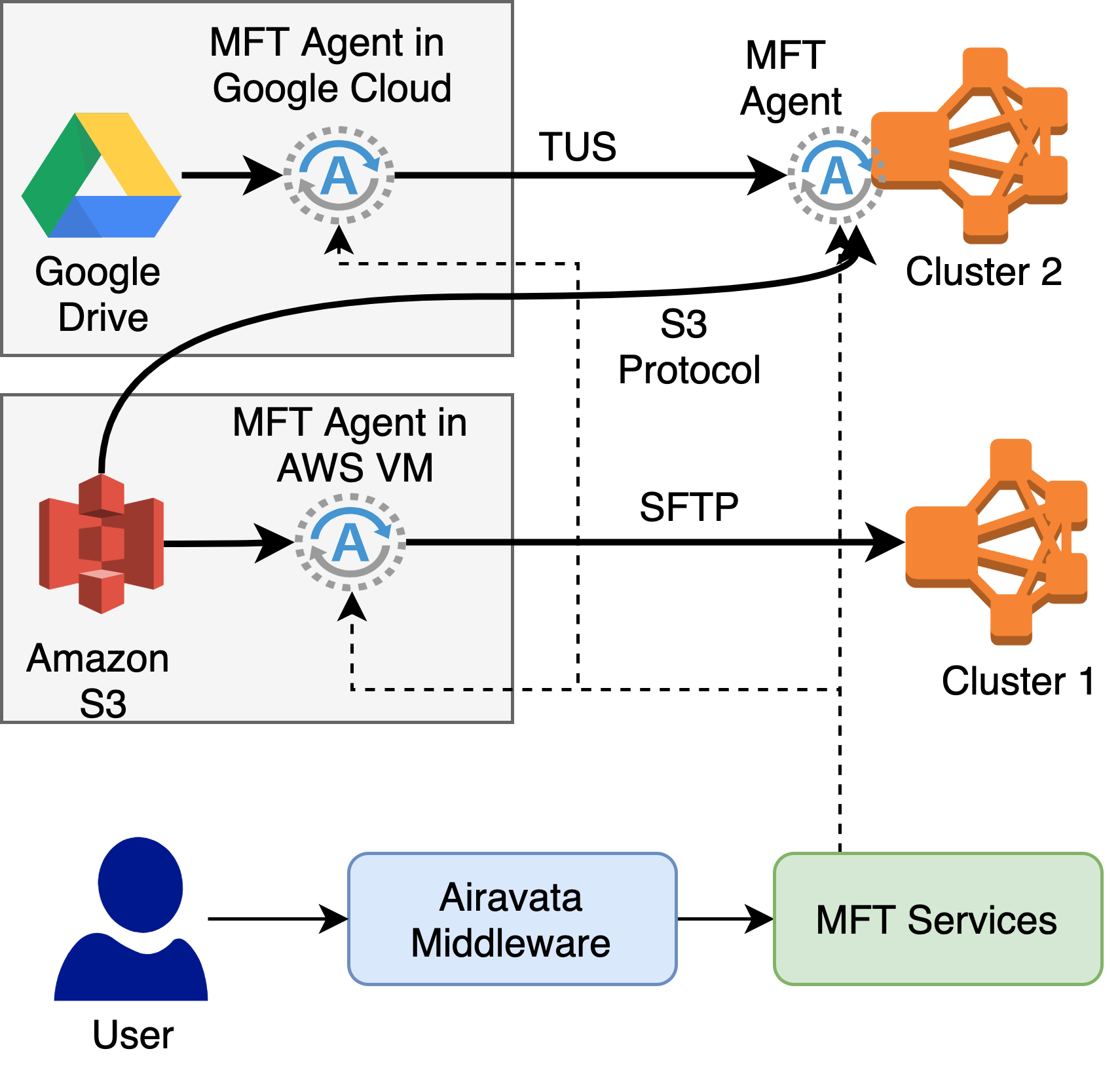}
  \caption{Handling cloud storage integration into gateways using Airavata MFT. Solid arrows depict data transfer paths and dashed arrows depict transfer control paths.
}
  \Description{Data Resource Management Service}
  \label{figCloudStorage}
\end{figure}

\textbf{Data transfers between cloud storage and HPC clusters:} When integrating cloud data storage, we can either install MFT Agents near the cloud storage or use an existing agent installed on a cluster to pull or push data using cloud storage supported protocol (Fig. \ref{figCloudStorage}). When installing a cloud MFT agent, we should make sure that the agent is installed in a place inside the same network subnet of the cloud provider. For example, S3 cloud agents can be installed within an AWS virtual machine. This approach provides the maximum transfer throughput with added network-level security between the cloud storage and the MFT Agent. 

\section{Related Work}
From a related work perspective, the capabilities discussed in this paper can be separated into storage management and data movement. There are multiple related efforts that cater to subsets of these capabilities. For instance, storage solutions such as MinIO \cite{minio} and Ceph \cite{weil2006ceph} provide a unified object storage abstraction mapping multiple file systems. Globus \cite{foster2011globus} is well known for scientific data management, particularly in the transfer of very large data sets. Globus data transfer services utilize GridFTP, a high performance, point-to-point data transfer protocol. The main advantage of this protocol is that the data and control paths are separated. This way, data can be transferred directly from one point to another while keeping the control path for another third party. Globus software is closed source, and its operations are proprietary. Open source projects like StorkCloud \cite{ross2014managed} and Rclone \cite{rclone} are MFT implementations that provide extensible multi-protocol transfer job schedulers and directory listing services. However, the data and control paths are interleaved in these approaches; separation of data and control paths is key for science gateway use cases. Alluxio \cite{alluxio} is a data orchestration framework that is heavily used in data analytics applications. Alluxio is a fully contained distributed system with its own data catalog and credential management system, making it challenging to integrate into other systems.

\section{Conclusion and Future Work}

In this paper we discussed our approach for diversifying data management for science gateways. The paper focuses on specific use cases that decouple data movement from both gateway web servers and middleware to enable direct transfers between users’ systems and computational resources. The capabilities discussed in this paper lay the foundation for empowering gateway frameworks to support diverse data-intensive computing such as more ubiquitous integration of cloud storage systems and support of data distribution gateways for scientific instruments. 

\begin{acks}
The Custos Integration is funded by the National Science Foundation Award number 1840003.
\end{acks}

\bibliographystyle{ACM-Reference-Format}
\bibliography{pearc21-mft-custos}

\end{document}